\documentclass[10pt,journal,compsoc]{IEEEtran}

\ifCLASSOPTIONcompsoc
  \usepackage[nocompress]{cite}
\else
  \usepackage{cite}
\fi
\ifCLASSINFOpdf
  \usepackage[pdftex]{graphicx}
  \usepackage{svg}
  \graphicspath{{img/}}
  \DeclareGraphicsExtensions{.pdf,.png,.svg}
\else
\fi
\usepackage{amssymb}

\usepackage{hyperref}

\usepackage{fixmetodonotes}

\usepackage{layouts}

\usepackage{amsmath}
\DeclareMathOperator{\Var}{Var}

\begin{document}
\title{Sensitive vPSA - Exploring Sensitivity\\ in Visual Parameter Space Analysis}

\newcommand{\toolname}{Sensitivity Explorer}

\author{Bernhard~Fröhler, 
        Tim~Elberfeld,
        Torsten~Möller,
        Hans-Christian~Hege,
        Julia~Maurer, 
        Christoph~Heinzl
\IEEEcompsocitemizethanks{\IEEEcompsocthanksitem B. Fröhler, J. Maurer and C. Heinzl are with the Research Group X-Ray Computed Tomography, University of Applied Sciences Upper Austria, Austria.\protect\\
E-mail: bernhard.froehler|julia.maurer|christoph.heinzl@fh-wels.at
\IEEEcompsocthanksitem T. Elberfeld is with the imec-Vision Lab, University of Antwerp, Belgium.
\IEEEcompsocthanksitem T. Möller is with Data Science @ Uni Vienna and the Faculty of Computer Science, University of Vienna, Austria.
\IEEEcompsocthanksitem H.-C. Hege is with the Visual and Data-Centric Computing Department, Zuse Institut Berlin, Germany.}
\thanks{Manuscript submitted \today}}

\ifCLASSOPTIONpeerreview
\markboth{Journal of \LaTeX\ Class Files,~Vol.~14, No.~8, August~2015}%
{Shell \MakeLowercase{\textit{et al.}}: Bare Demo of IEEEtran.cls for Computer Society Journals}
\else
\markboth{Submission x}%
{test}
\fi


\IEEEtitleabstractindextext{%
\begin{abstract}
The sensitivity of parameters in computational science problems is difficult to assess, especially for algorithms with multiple input parameters and diverse outputs.
This work seeks to explore sensitivity analysis in the visualization domain, introducing novel techniques for respective visual analyses of parameter sensitivity in multi-dimensional algorithms.
First, the sensitivity analysis background is revisited, highlighting the definition of sensitivity analysis and approaches analyzing global and local sensitivity as well as the differences of sensitivity analysis to the more common uncertainty analysis. 
We introduce and explore parameter sensitivity using visualization techniques from overviews to details on demand, covering the analysis of all aspects of sensitivity in a prototypical implementation.
The respective visualization techniques outline the algorithmic in- and outputs including indications, on how sensitive an input is with regard to the outputs. The detailed sensitivity information is communicated through constellation plots for the exploration of input and output spaces. A matrix view is discussed for localized information on the sensitivity of specific outputs to specific inputs. A 3D view provides the link of the parameter sensitivity to the spatial domain, in which the results of the multi-dimensional algorithms are embedded.
The proposed sensitivity analysis techniques are implemented and evaluated in a prototype called \toolname{}.
We show that \toolname{} reliably identifies the most influential parameters and provides insights into which of the output characteristics these affect as well as to which extent.
\end{abstract}

}

\maketitle

\IEEEdisplaynontitleabstractindextext

\IEEEpeerreviewmaketitle

\IEEEraisesectionheading{\section{Introduction}\label{sec:introduction}}

\IEEEPARstart{T}{he} analysis of the sensitivity of a computational model, its inputs and outputs, is important in many application domains of scientific computing.
Saltelli et al. define sensitivity analysis (SA) as
``The study of how the uncertainty in the output of a model (numerical or otherwise) can be apportioned to different sources of uncertainty in the model input'' \cite{Saltelli:2004}.
As an effect of this rather narrow definition, in current visualization research sensitivity analysis is most often revisited in the context of uncertainty analysis. 
This definition is appropriate in cases, where the input (and therefore also output) parameters are random.
The sensitivities tell us here, how the input uncertainty is mapped (in a linear approximation) to the output uncertainty. 
However, sensitivities are also of interest when no randomness is involved: 
The sensitivity tells us in that case, which parameters influence the result in a particularly strong way, and which of the parameters have almost no influence on the result.
A broader definition of sensitivity analysis, which also fits to this case, is found in the following description:
``Sensitivity analysis is any study of variation in the output of some system or some algorithm with respect to changes in the input.''

Analyzing sensitivity in this broader sense of detaching sensitivity from uncertainty analysis, has received relatively little attention so far in visualization research.
When analyzing the parameter space of image analysis methods for reconstructing characteristics of fibers in fiber-reinforced polymers,
we realized the need to know more about the behavior of the algorithm.
Our collaborators, which are developers of fiber reconstruction algorithms as well as users of such algorithms,
were mainly interested in the detailed analysis of the effects of changes of input parameters on the output(s).
That is, how the change of a single parameter propagates through the analysis framework and how it influences the output characteristics. 
On the one hand, global sensitivity is important in this respect. Global sensitivity analysis provides information on which input parameters have the largest effect on the various output characteristics.
On the other hand, local sensitivity is of equal importance in order to explore a punctual impact of an input parameter on the output(s). 
We prefer, however, to consider \emph{regional} sensitivity, i.e., the effects of small perturbations in multiple different locations in parameter space. 
To create a comprehensive understanding, we designed a visual interface for analyzing sensitivity in a computational setting. 
We tested our proposed techniques in terms of sensitivity analysis of fiber reconstruction algorithms. 
The main contributions of this work are thus found in the following aspects:
\begin{itemize}
    \item a formal definition of sensitivity analysis for computational models, distinguishing between global, regional, and local analysis
    \item a clarification of sensitivity analysis in contrast to uncertainty analysis
    \item methods for performing sensitivity analysis for algorithms producing complex data, where the classical mathematical definitions fail
    \item an implementation of these methods for the analysis of fiber reconstruction algorithms in the prototype \toolname{} 
\end{itemize}

We start by exploring related work on sensitivity analysis in \autoref{sec:background}.
In \autoref{sec:methods}, we present a general analysis framework for sensitivity in algorithms with complex output. 
\autoref{sec:application} discusses our concrete implementation of this framework for fiber reconstruction algorithms along with an introductory example presented with synthetic data.
Finally, a limited discussion of our results and future work can be found in \autoref{sec:discussion}.

\section{Background}
\label{sec:background}


Sensitivity analysis (SA) is typically categorized into the analysis of local and global sensitivity.
Local sensitivity measures only the variation at a fixed location in the parameter space,
while global sensitivity measures the variation in a wide parameter range.
In the earlier works of Saltelli et al. \cite{Saltelli:2004, Saltelli:2008}, an introduction into SA has been provided.
More recently, an overview on the advances in the field was presented by Borgonovo and Plischke \cite{Borgonovo:2016}, together with extensive overviews on how SA is applied in different research areas as provided by Saltelli et al. \cite{saltelli_why_2019}.
In the latter work, the authors reason that the most prevalent one-at-a-time analysis,
where one parameter at a time is varied around a fixed position, is only useful if there is a linear correlation between input and output.
They argue that more global methods should be used for solid SA whenever such a linear correlation cannot be ascertained.
Furthermore, Razavi and Gupta \cite{Razavi:2015} identify two major challenges associated with global SA (GSA):
First, multiple differing global SA methods do exist, but they are based on different philosophies and theoretical definitions of sensitivity.
Second, there is a high computational cost to carrying out global SA, which often leads to local, rather superficial SA being performed.
The authors address these shortcomings with their variogram-based SA methods \cite{Razavi:2016a, Razavi:2016b} implemented in the VARS-TOOL \cite{Razavi:2019}.
They achieve an efficient SA by employing similar sampling strategies as Berger et al. \cite{Berger:2011}.
The focus of Berger et al. however is on enabling users to quickly get an impression of the parameters space, also in locations between the sampling points, when the algorithm to be investigated is slow and the user would need to wait for the result.
They suggest to use a surrogate model to speed up the analysis.






In terms of visualization of global SA, there is surprisingly little literature to be found on this subject yet.
Popelin and Ioosss are amongst the first to explicitly address visualization methods for SA \cite{Popelin:2013}.
They analyze the variation of curves spanned by several output variables, such as temperature or pressure, over time in a nuclear reactor simulation.
Their methods aim at finding the average curve, a confidence interval containing most curves and outliers.
Hence, this work is specific to the analysis of temporal data.
Torsney-Weir et al. \cite{TorsneyWeir:2018} performed a user study on whether
and how visualization of sensitivity affects the behavior of users in an investment scenario.
They found their users can be categorized into two groups: one which utilizes the sensitivity in their analysis and trusts the system more if present, and another group which tends to ignore the provided sensitivity information.
Pöthkow and Hege \cite{Poethkow:2010, Poethkow:2012} introduced methods for exploring sensitivity and resulting uncertainty in terms of iso-contours and surfaces, for the analysis of a fuel injection processes and weather simulations.
A few other works touch on SA briefly but do not put their main focus on it.
Hazarika et al. \cite{Hazarika:2019:NNVA}, for example, provide a tool for the visual analysis of uncertainty and sensitivity of a yeast cell polarization simulation.
In contrast to this work, we do not provide a surrogate model, and our current analysis showed no need for refining our sampling. Our focus is more on the exploratory aspect, and we visualize a much broader range of information related to sensitivity.

In the context of image analysis,
Torsney-Weir et al. provide some support for SA in their Tuner tool \cite{Torsney-Weir:2011:Tuner}, applied on the comparison of segmentation methods.
In their work, SA is only a side aspect, and they require a ground truth for the analysis of the segmentation quality.
Cortez and Embrechts~\cite{Cortez:2013} proposed SA methods and visualizations for opening black box supervised learning models. 
The PorosityAnalyzer tool by Weissenböck et al. \cite{Weissenboeck:2016:PorosityAnalyzer} provides implicit SA by showing the correlations between input parameters and output measures in a scatter plot matrix. 
Previous work by Fröhler et al. \cite{Froehler:2018:FIAKER, Froehler:2020:Curved} for the analysis fiber reconstruction algorithms focused on analyzing and comparing single results.
In line with addressing the quantitative data visualization challenge laid out by Heinzl and Stappen \cite{Heinzl:2017:STARVisMat},
we focus on a quantitative view of the sensitivity,
and provide feedback on the consequences of parameter changes across the parameter space.
The work of Brecheisen et al. \cite{Brecheisen:2009} investigates the the sensitivity of parameters to an algorithm analyzing diffusion tensor imaging.
From a conceptional point of view theirs is the closest to our work which we have found.
Whereas they provide means for a qualitative, spatial exploration of variations in up to a few fibers, analyzed across a few algorithmic parameters, 
we enable the quantitative analysis of many parameters and many fibers at once.


\section{Methods}
\label{sec:methods}
To explain and introduce our approach for analyzing the sensitivity of a computational model, we start by describing what we mean by ''sensitivity analysis of empirical functions''.
Our focus lies on image analysis algorithms as such empirical functions. Our terminology in describing respective functions or algorithms is based on the conceptual framework for visual parameter space analysis by Sedlmair et al. \cite{sedlmair_visual_2014}.
We propose a multi-step analysis workflow for analyzing the sensitivity of a multi-dimensional empirical function.
It is depicted in \autoref{fig:workflow}, illustrated on the example of the analysis of tomographic reconstruction algorithms.
First, the input parameter space of the analyzed algorithm must be sampled and the corresponding outputs calculated.
The sampling is followed by the computation of sensitivity measures,
which then in turn can be analyzed through the proposed visual analysis methods.

\subsection{Sensitivity analysis for empirical functions}

Sensitivity analysis is a topic of research in the visualization community and still to a large extent unexplored. Currently, it is often mixed with uncertainty analysis or seen as a sub-area of it. With this chapter, we intend to give a clear understanding of sensitivity analysis, how it differs from uncertainty analysis, and why it is important to create visual tools supporting sensitivity analysis. Because of the frequent misunderstanding of seeing sensitivity analysis as just a part of uncertainty analysis, we have deliberately chosen a didactic approach.

\subsubsection{A one-dimensional example}

\begin{figure}[ht]
    \centering
    \includegraphics[width=\columnwidth]{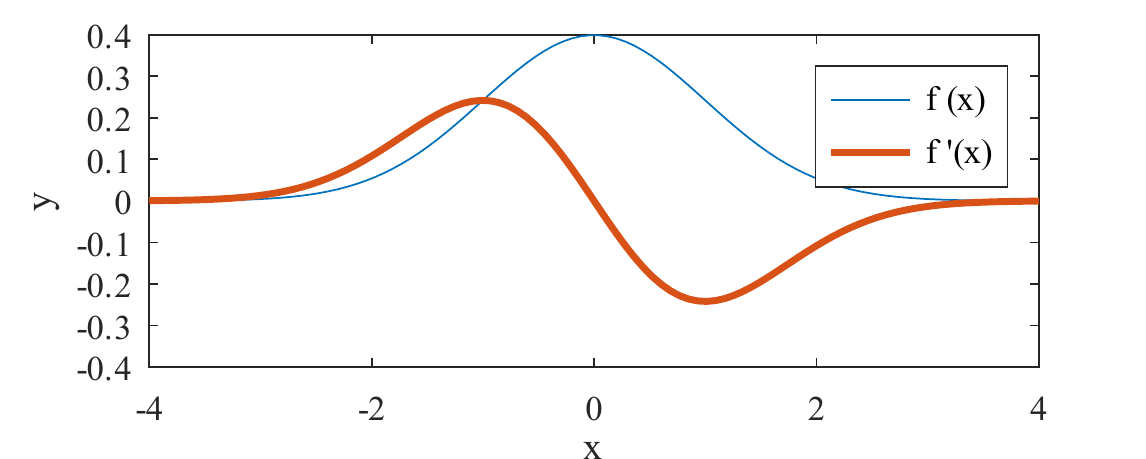}
    \caption{Gaussian function from \autoref{equ:gaussian} ($\mu=0, \sigma=1$) and its derivative from \autoref{equ:gaussDeriv}}
    \label{fig:gaussfunctions}
\end{figure}

Let's assume we are given a system with an input variable $x$ and an output variable $y=f(x)$, described by a 1D function $f:\mathbb{R} \rightarrow \mathbb{R}$, e.g.\ a Gaussian function
\begin{equation}
\label{equ:gaussian}
f(x) = \frac{1}{\sigma\sqrt{2\pi}}\mathrm{e}^{-\frac{(x-\mu)^2}{2\sigma^2}}.
\end{equation}
The \emph{sensitivity} of this system at a particular value $x$ can be simply described as how much the output $f(x)$ changes when its input $x$ changes. Mathematically speaking, this is given by the first order derivative $f'(x)$ of that function, i.e.\ in the case of a Gaussian, a ``DoG'' function:
\begin{equation}
\label{equ:gaussDeriv}
f'(x) = \frac{-x}{\sigma^3\sqrt{2\pi}}\mathrm{e}^{-\frac{(x-\mu)^2}{2\sigma^2}}.
\end{equation}
A plot of the function $f$ and its derivative $f'$ can be seen in \autoref{fig:gaussfunctions}. 

The derivative $f'(a)$ provides the sensitivity of $f$ at point $a$. Where is system most sensitive to change? At the edges of the function $f$, because there a small change in $x$ causes a large change in $f(x)$.

Hence, in no sense of the word ``uncertainty'' is there anything uncertain: we know everything about the function; it is a deterministic function. 
Since this sensitivity is measured at a single point of the input, it is called \emph{local sensitivity}. 

However, the \emph{global sensitivity} of a function measures the variance of this function over a particular parameter range. That is, one estimates, how much the value of a function varies over a set of input values 

\begin{equation}
    \Var_f([-t,t])=\int_{-t}^t (x-E(x))^2f(x)\,dx
\end{equation}
where $E(x)$ is the expected value, or the mean, of the function in the integration domain. In case of the standardized Gaussian, for large $t$ the mean is $\mu$ and the variance is $\sigma^2$; if the interval $[-t,t]$ is narrow, some adjustments are needed.

While our didactic example is done in 1D only, local as well as global sensitivity is naturally extended to multiple dimensions. 
In local sensitivity, partial derivatives are then used instead of the derivative, i.e.\ changes in the function are looked at for small changes in one of the input variables.
More generally, one can consider the derivative along any given direction vector, i.e.\ linear combinations of partial derivatives.
And for global sensitivity, one is integrating over an interval, typically in one input variable, keeping all other variables fixed. The integration interval is defined to contain the values typically occurring in the application under consideration. Using such global sensitivities, the relative influence of each input variable on the total variance can be determined. 

Now what is the relation to uncertainty analysis? First, let's look at 1D  again. Let's assume the input value $x$ is subject to some uncertainty, e.g.\ that it lies in some interval $[x,x+ \Delta x]$. Then the question arises how this uncertainty propagates to an uncertainty of the output variable $y=f(x)$. According to the rules of analysis, a differentiable function $f$ can be well approximated in the interval $[x, x+ \Delta x]$ for sufficiently small $\Delta x$ by a linear function $f(x) + f'(x)\, \Delta x \approx f(x+\Delta x) = y+\Delta y$, i.e.\ $f'(x)\, \Delta x \approx \Delta y$. That is, the local sensitivity $f'(x)$ represents the \emph{amplification or damping factor} with which the uncertainty $\Delta x$ is propagated to uncertainty $\Delta y$. Therefore, there is a close connection between local sensitivities and uncertainty propagation.

Again, this can be directly applied to the multidimensional case: the graph of a multivariate function $f:\mathbb{R}^m \rightarrow \mathbb{R}$ is locally approximated by the tangent plane. And the constants of this linear function are the partial derivatives, i.e.\ the local sensitivities, which together form the gradient $\nabla f$ at the considered point. The linear relation becomes $\nabla f(x)\cdot \Delta x \approx \Delta y$, where $\Delta x$ is a vector now.
And in case of a multivariate vector function $f:\mathbb{R}^m \rightarrow \mathbb{R}^n$ the constants of the approximating linear function are the partial derivatives, i.e.\ the local sensitivities, comprising the Jacobian matrix $J(f)$ at the considered point. The linear relation describing the uncertainty propagation from input variables $x$ to output variables $y$ becomes $J(f)(x)\Delta x \approx \Delta y$, where $\Delta x$ and $\Delta y$ are vectors now.

It is important to note that for systems described by a deterministic function $f$, the sensitivities are deterministic and have a value in themselves -- independent of any uncertainty considerations.

When the system under consideration is itself stochastic, i.e.\ when $f$ becomes a random function, the sensitivities also become random variables and are afflicted with uncertainty themselves -- but that is another story, which we do not consider here.

\begin{figure}
    \centering
    \includegraphics[width=\linewidth]{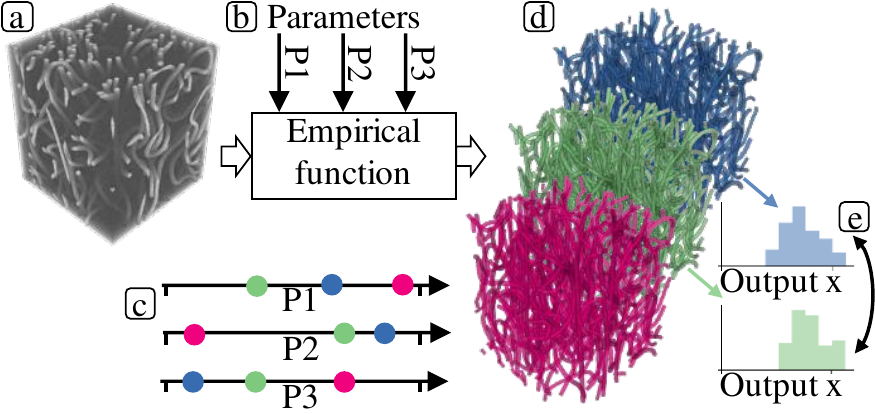}
    \caption{Workflow in computing sensitivity information: A complex object (a) together with some parameters (b) are input to an empirical function; sampling over the parameter space (c) yields complex objects as output (d), for which the variance is computed, for instance from the distribution of a specific output characteristic (e).}
    \label{fig:workflow}
\end{figure}


\subsection{Sensitivity sampling}
\label{sec:sampling}
To analyse the sensitivity of any given pipeline, first a systematic sampling of its input parameter space needs to be performed, followed by computing the respective output of each sampled parameter set.
To cover the full parameter space indicated by the user, Latin Hypercube sampling is used.
The desired sample size $N$ is currently a parameter to be determined by the user.
For each parameter set obtained from this initial sampling,
we perform an additional star-shaped sampling to compute the local sensitivity, similar to the work by Berger et al. \cite{Berger:2011} and Razavi et al. \cite{Razavi:2016b}.
For each such \emph{star}, we create a star \emph{branch} with additional parameter sets for each input parameter,
where this parameter is varied according to a fixed step width $w$,
and all other parameters are kept fixed at the original value.
An exemplary fiber reconstruction algorithm with three input parameters, along with three outcomes resulting from different parametrizations is visualized in \autoref{fig:workflow}(a). 
The step width $w$ is determined by the user as a fraction of the total range covered by a parameter.
The analyzed algorithm is subsequently executed for each parameter set.

\subsection{Sensitivity measures}
\label{sec:sensitivity-measures}


Based on the sampling as described above, we compute the local and global sensitivity of each input parameter with respect to each output.
Image analysis pipelines can be analyzed in the terminology of Sedlmair et al.'s \cite{sedlmair_visual_2014} taxonomy for visual parameter space analysis.
Such pipelines typically take numerical parameters as input, along with one or more complex objects.
As output, they deliver one or more complex objects.
For our exemplary fiber reconstruction pipeline this is depicted in \autoref{fig:workflow}.
The complex objects in the input of these pipelines take the form of three-dimensional volumes, resulting for example from computed tomography imaging.
The complex objects in the output consist of a collection of fiber objects, each described by a number of characteristics such as length, diameter, volume, surface area, or orientation.
For sensitivity analysis, no direct objective measure for the quality of a specific output is required.
Instead, derived measures only need to quantify differences in the outputs resulting from varying input parameters.
We propose two types of derived measures:
\begin{itemize}
    \item Measures based on the differences in the distribution of object characteristics
    \item Measures based on the pairwise, best-match differences between two objects
\end{itemize}
Each of these measures is a function $m(a, b) \to \mathbb{R}$ computing the difference between two outputs $a$, $b$. 
Each output is represented by a set of objects $O^a=\{o^a_1, o^a_2, ... o^a_m\}$, $O^b=\{o^b_1, o^b_2, ... o^b_n\}$, with $n$ and $m$ denoting the number of objects in $a$ and $b$, respectively. 

\textbf{Distribution-based measures} are rooted in the observation that domain experts are typically interested in the exact distributions of the object characteristics.
They are therefore interested in the influence of a parameter on the distribution of specific output characteristics,
we experimented with two difference measures for distributions, represented by their histograms:
One approach is to compute the mean Euclidean distance between the histograms interpreted as vectors in a space with as many dimensions as there are bins in the histogram.
As a more statistically grounded second approach, we employ the Jensen-Shannon divergence \cite{Endres:2003:NewMetric}, which is a measure covering the information radius or the total divergence to the average.
For each pair of outputs $a$, $b$, this results in a difference value for each characteristics distribution under investigation.
As an intermediate step for computing the full distribution difference, we also compute the variation in each single histogram bin.
This provides additional information into where in the distribution changes are most likely to occur, as will be shown in \autoref{sec:application}.

\textbf{Pairwise, best-match-based measures} are computed in multiple steps.
We first compute the object dissimilarity $s(o^a_i, o^b_j)$ for each pair $(o^a_i, o^b_j) \in O^a \times O^b$, 
resulting in a matrix of the pairwise dissimilarities of the objects in $a$ to the objects in $b$.
For each object $o^a_i$ in $a$, we then chose the object $o^b_j$ in $b$ with the lowest dissimilarity as best match.
The next goal is to compute a pairwise difference $p(o^a_i, o^b_j)$ for each such best-match pair.
Depending on the specific use case, a number of difference measures are employed.
The dissimilarity measure used above to determine best matches can be directly applied.
Additional measures based on the difference in single object characteristics are employed to analyze the variation specific to these characteristics.
The overall difference between two outputs $a$ and $b$ is then determined for each defined difference measure as the average over the pairwise differences of all best-matching fiber pairs.
For a practical example of such a measure, see \autoref{sec:fiberDifferenceMeasures}.

Both types of difference measures as defined and explained in the previous paragraphs provide a way to measure variation in the output of the investigated pipeline.
Sensitivity based on these measures is computed from neighbouring parameter sets in the parameter space.
For each parameter set from the initial sampling, the local sensitivity is computed as the mean of the variation between the initial parameter set
and its neighbours in the star sampling branch for this parameter.
As discussed, multiple separate measures are considered, e.g., for each output characteristic.
Therefore, we also arrive at multiple local sensitivity measures, typically one per output characteristic, and additional ones per available dissimilarity measure.
Global sensitivity measures are computed by averaging the local sensitivity measures.
These measures enable thorough insights into the variation caused by parameters, 
as well as the influence of specific parameters on specific characteristic, 
which will be demonstrated at the example of our prototypical implementation for the sensitivity analysis of fiber reconstruction.

\section{\toolname{} Prototype}
\label{sec:application}
This section describes our prototype for sensitivity analysis (SA) on the example of fiber reconstruction algorithms, named \toolname{}.
Fiber reconstruction algorithms are used to directly extract and characterize features, that is, spatial objects such as fiber-reinforced polymers, from X-ray computed tomography (XCT) data.
\toolname{} analyzes a collection of results sampled according to the procedure as described in \autoref{sec:sampling}.
When loading the collection, sensitivity measures are computed at first as described in \autoref{sec:sensitivity-measures}, with adaptations for the specifics of fiber reconstruction results as in the following \autoref{sec:fiberDifferenceMeasures}.

\subsection{Output difference measures} 
\label{sec:fiberDifferenceMeasures}
Given two results computed from the same input with different parameterizations or algorithms,
we need a way to determine whether the same fibers were found,
and how similar each pair of matching fibers in the two results is, as described in \autoref{sec:sensitivity-measures}.
Fröhler et al. proposed a number of dissimilarity measures for fiber reconstruction results \cite{Froehler:2018:FIAKER, Froehler:2020:Curved}.
These measure the dissimilarity between two spatial fiber objects,
either based on the similarity of their output characteristics (such as position, length, orientation, and diameter),
the distances between the points of which these fibers consist,
or the ratio of volume overlap.

If two fibers feature exactly the same characteristics, these normalized measures deliver a value of 0, and respectively higher values up to 1 for more dissimilar fibers.
As determined by Fröhler et al., the overlap-based measure fits best to the expectations of users regarding the evaluation of fiber reconstruction results considering that the output results from the same input.
The first step is to sample a fixed number of points in fiber $f_x$, and to check the containment of each point in the other fiber $f_y$. The result is then computed as one minus the ratio of points contained within $f_y$ divided by the number of sampled points, with an additional correction factor determined by the ratio of the volumes of $f_x$ and $f_y$.


We consider each pair of reconstruction results $a$ and $b$, and apply this measure for each fiber $f^a_i$ in result $a$ to find the fiber $f^b_j$ in result $b$ with the lowest dissimilarity (as mentioned in \autoref{sec:sensitivity-measures}, we call $f^b_j$ the \emph{best match} of $f^a_i$).
A brute-force approach to find the best match for $f^a_i$ is to compute the dissimilarity to all fibers in $b$; the best match is then simply the fiber with the lowest dissimilarity.
However, calculating the dissimilarity between two fibers is rather computation-intensive.
We therefore applied an optimization using bounding boxes: In a preprocessing step before the dissimilarity computation, the bounding boxes of all fibers in all results are determined. The dissimilarity measure is then only calculated for fiber pairs with overlapping bounding boxes.
If there is no fiber in result $b$ with overlapping bounding boxes, the dissimilarity to $b$ for $f^a_i$ is set to 1.
As mentioned in \autoref{sec:sensitivity-measures}, an aggregated dissimilarity between the two results can then be computed as the mean of the dissimilarity of these best matches.

\subsection{Visual sensitivity analysis in \toolname{}}

\begin{figure}[tb]
    \centering
    \includegraphics{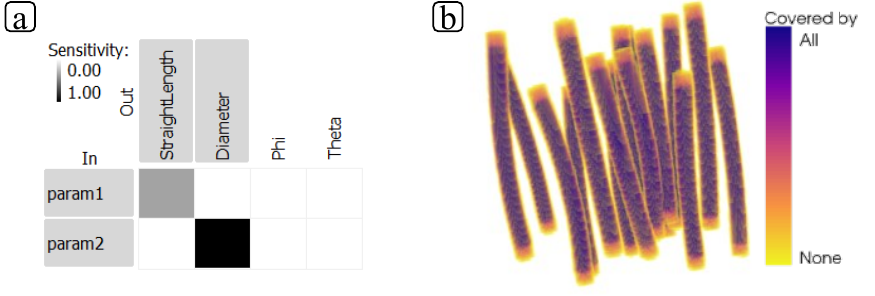}
    \caption{The \emph{in-out matrix} (a) provides an overview of the influence of input parameters (rows) on output characteristics (columns). The spatial overview (b) visualizes the occupation ratio of the volume across results.}
    \label{fig:overview}
\end{figure}

In parallel to elaborating the sensitivity analysis considerations laid out in \autoref{sec:methods},
we implemented a prototype for the analysis of sensitivity in fiber reconstruction pipelines
in order to practically test our methods.
The \toolname{} prototype which we present here follows the information seeking mantra by Shneiderman \cite{shneiderman_eyes_1996}.
It consists of multiple views linked by a variety of interactions to provide detai whenever necessary.

\subsubsection{\toolname{}'s views}

The \emph{in-out matrix} (see \autoref{fig:overview}(a)) provides an overview on available input parameters of the fiber reconstruction pipeline in the rows, and the available output characteristics in the columns.
The matrix cells are colored according to the correlation of the respective input parameter and output characteristic, as measured by the respective global sensitivity, resulting in a heat map: High sensitivity is encoded in black, whereas lower sensitivity values fade towards white.

The \emph{spatial view} (see \autoref{fig:overview}(b)) initially provides a cumulative visualization of the variation across all results.
It is computed by projecting single fibers from all results to voxel space, and counting how many fibers cover a single voxel. 
Divided by the number of results, this typically results in a measure from 0 to 1 (with potential values above 1 if the voxel resolution is chosen so low that some voxels are touched by multiple fibers even if those fibers don't directly overlap). So in the spatial view areas covered by all results are rendered in dark blue, whereas areas covered by a single object only are colored yellow; areas not covered by any fiber remain white. 
When one or more results are selected, their model-based representations are displayed instead.

\begin{figure}[tb]
    \centering
    \includegraphics[width=.95\linewidth]{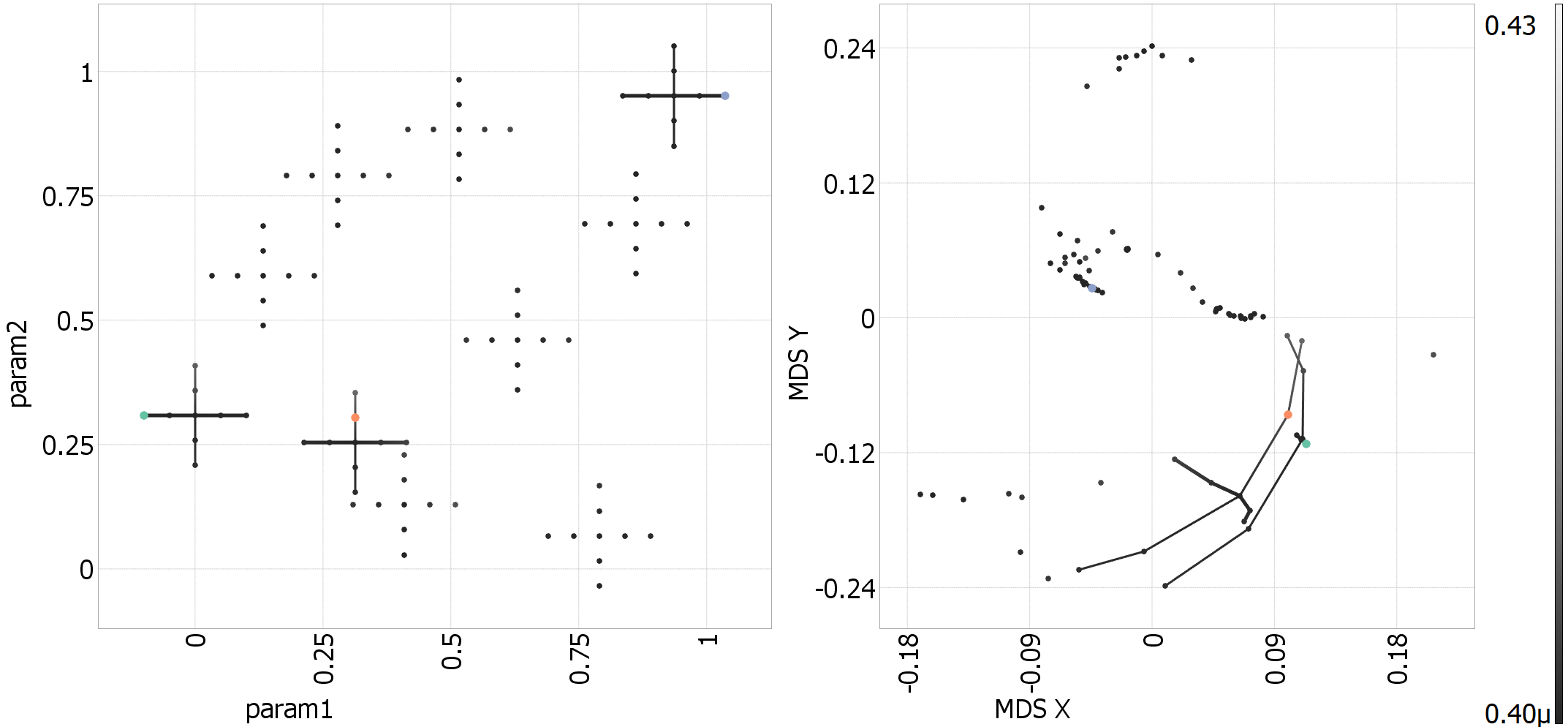}
    \caption{\emph{Constellation plots} show the results in scatter plots, with lines visualizing the sampled star branches in the parameter space (left), and in an MDS embedding based on the average result dissimilarities (right).}
    \label{fig:constellation}
\end{figure}

The \emph{constellation plots} (see \autoref{fig:constellation}), which we named for their resemblance to astronomical constellations, show how fiber reconstruction results are distributed and connected through their sampling.

Constellation plots consist of two views:
The \emph{parameter space plot} on the left displays the distribution of results in the parameter space, that is, it visualizes the sample points.
A multi-dimensional scaling (MDS) embedding of the results is shown in the \emph{MDS plot} on the right.
An MDS embedding shows items mapped to a lower dimension, while trying to represent their given pairwise distances as accurately as possible \cite{Bronstein:2009}.
In our case, the distances between results are computed according to their average dissimilarity, as described in \autoref{sec:fiberDifferenceMeasures}.
Both scatter plots display connecting lines for the sampled stars in the parameter space plot of which one or more members are selected.
The single dots are color-coded by dissimilarity: If no result is selected, from the star center, otherwise, the dissimilarity of the points is color-coded with reference to the result selected last.
By default, a gray scale color map is used here.
For the coloration of the constellation lines, a linear gradient between the color of the two connected dots is applied.
The thickness of the constellation line differs in 3 steps, and encodes the parameter branch the line visualizes:
The lines following the parameter branch for the parameter shown on the x-axis of the parameter space plot are thickest,
followed by those for the parameter on the y-axis of the parameter space plot. 
All other parameter branches, that is, those currently not shown in the parameter space plot, are only shown using very thin lines.
By default, the two parameters with the highest global sensitivity are shown on the x- and y-axis of the the parameter space plot.

\begin{figure}[tb]
    \centering
    \includegraphics{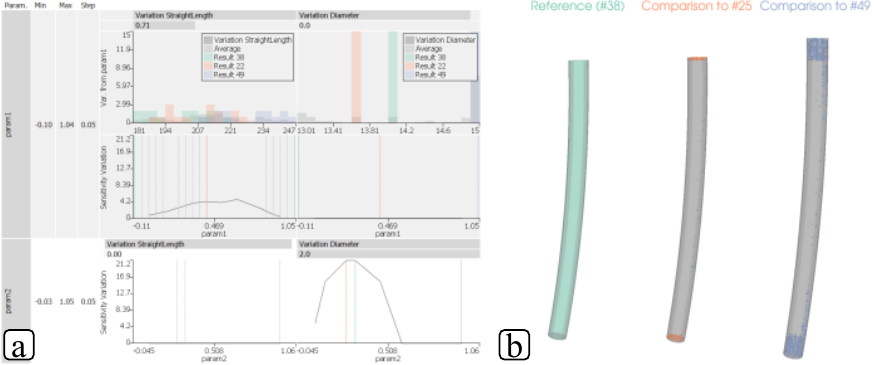}
    \caption{Detailed information on the influence of input parameters on selected output characteristics is shown in the \emph{parameter influence view} (a). The \emph{fiber difference view} shows small multiples for the differences between single fibers of selected results (b).}
    \label{fig:detail}
\end{figure}

The \emph{parameter influence view} (see \autoref{fig:detail}(a)) provides details on demand on particular combinations of input parameter and output characteristic, as selected in the in-out matrix or in the constellation plot by selecting a result on a specific parameter branch.
It uses the same matrix layout and ordering as the in-out matrix, but its cells provide two detail charts.
First, it displays a histogram of the average distribution of the selected characteristic,
along with a plot of the variation for each bin of the histogram.
Second, it shows a regional sensitivity plot, that is, the local sensitivity over the investigated parameter range.
The parameters can be sorted by their influence on output characteristics, as measured by the global sensitivity for that characteristic.
This affects the in-out-matrix and the parameter influence view.
By default, they are sorted by their influence on the average length of the fibers.
The parameter values of currently selected results are shown with colored markers, along with light-gray markers indicating other results on the same star branch.

On selecting one or more fibers as well as two or more results, the \emph{fiber difference view} (see \autoref{fig:detail}(b)) provides details on the differences to the best matches between these results.
The first selected result is shown in the leftmost view, as reference. 
For each additionally selected result, the difference to the reference is explicitly encoded:
The original fiber from the reference is shown in gray.
Areas dotted in the color of one of the two results indicate regions that are only covered by the fiber in the respective result.
The image in the very left of \autoref{fig:detail}(b)), e.g., indicates that the fibers in \#49 are considerably shorter than their best matches in the reference,
as indicated by the green dotted regions above and below each fiber.
A fiber selection triggered by the user is always kept as active fiber selection until the user either cancels the selection or selects a new set of fibers.
For each newly selected result, the best match to the active selection is shown, and compared to the reference.



\begin{figure}[htb]
    \centering
    \includegraphics{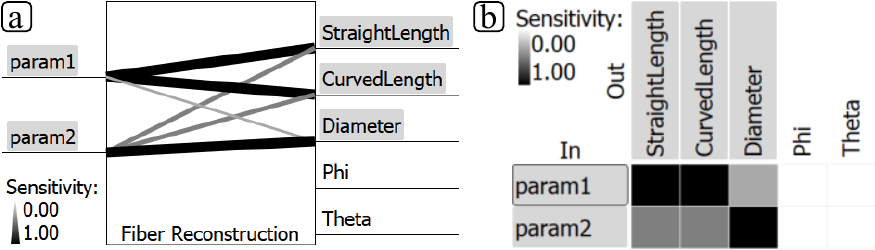}
    \caption{Explored options for detail on algorithm input and output: Box view with connecting lines between inputs and outputs encoding the correlation (a), and a matrix in vs. out view (b).}
    \label{fig:algorithmDetail}
\end{figure}

\subsubsection{Design decisions}

Our prototype was designed over several iterations and discussion rounds with domain and visualization experts.

For the overview over the input and output of an algorithm, and their correlation,
we were experimenting with two different designs.
The design presented in \autoref{fig:algorithmDetail}(a), with inputs on the left and outputs on the right, more clearly communicates the flow of information through the fiber reconstruction pipeline.
The display of relations between input and output through lines with varying color (gray scale) and width, however, is only suitable for a very limited number of connections.
We therefore switched to a matrix-based view, which scales much better; it is also more consistent to the parameter influence view.

In the beginning we experimented with a result dissimilarity matrix with matrix cells colored according to the similarity of the two results at the intersection.
Results could be sorted according to parameter values.
This revealed some patterns, such as if a parameter triggered larger changes.
But these patterns made the desired information only visible indirectly:
A parameter with high influence on the outcome would be characterized
by a pattern of colors varying gradually from one side to the other over the full spectrum of the color scale.
Parameters with little influence were characterized by local patterns of high variation.
Also, the layout of the matrix didn't lend itself well to visualizing the parameter space,
as rows and columns had fixed sizes, which does not reflect the position of the respective result in the parameter space.

For providing details on specific combinations of input parameters and output characteristics, we soon established the tabular layout.
However, it turned out to be tricky to combine an overview over the global sensitivity of such a pair with the detailed information charts on regional sensitivity as well as characteristics distributions.
We experimented with a LineUp-like interface \cite{gratzl_lineup_2013},
but the combination of aligned bar charts with detail charts proved to be challenging.
Remnants of this experiment still remain in our parameter influence view,
where each cell still provides a bar chart for the global sensitivity (showing also the numeric value), along with a reference bar with the caption for this cell.
The quick overview over which parameter influences which output is now instead provided in the in-out matrix.
Also, showing detail charts for all input parameters and output characteristics at once would be too overwhelming.
So, inspired by LiveRAC by McLachlan et al. \cite{mclachlan_liverac_2008}, the parameter influence view expands only those cells which currently are selected for inspection in more detail.

We experimented a lot with how best to use colors in our elaborate multi-view system.
we initially used red and blue to distinguish between input parameters and output characteristics, respectively.
However, we abandoned this encoding since this became too confusing once we added colors for encoding other data.
Instead, we now use a consistent layout for both the in-out matrix and the parameter influence view.
We also experimented with different color maps in other places.
For results, we now use a qualitative color map from colorbrewer \cite{Harrower:colorbrewerorg:2003}.
This color is consistently applied to results selected in the constellation plots, for showing the distribution of a result in the parameter influence view, for showing results in the spatial view, as well as for encoding the changes between results in the fiber difference view.
The spatial overview is colored using a perceptually uniform color map, by default the matplotlib's \emph{Plasma} color map is used \cite{smith_better_2015}.
Apart from this, there are only two grayscale color maps in use: One is used in the in-out matrix to encode the sensitivity of an output characteristic against changes in an input parameter; the second encodes differences between results in the constellation plots.

\subsubsection{Implementation details}
All of our methods are implemented in C++ as part of the open\_iA tool \cite{Froehler:2019:JOSS}.
Our sampling methods are based on a tool originally created for segmentation algorithms \cite{Froehler:2016:GEMSE}.
It was extended so that it can be used for all image processing filters included in the open\_iA framework.
Alternatively, external pipelines can be triggered through a command line interface.

The analysis prototype is implemented as a separate module.
On its first run on a newly sampled collection of results, dissimilarity measures are computed.
For 130 results, each with an average of 77 fibers, the required preprocessing takes approximately 8 minutes 30 seconds on an AMD Ryzen 5900X with 32GB of RAM.
The vast majority of this time (approximately 4 minutes each) is spent in computing the pairwise dissimilarities as well as computing the spatial overview image.
The result of these computations is cached on disk to speed up subsequent analysis.
That is, the second time loading the same collection of results only takes a few seconds.
Full fiber datasets from computed tomography might be much larger and contain up to several hundreds of thousands of fibers.
In our experience, though, it is not very useful to investigate such a large dataset as a whole, since the large size only complicates focusing on its details.
It is typically enough to extract a smaller region of interest to gain meaningful insights on the behaviors of algorithms.

\subsection{Introductory example and verification }
To test the correctness of our approaches and to provide a simple introductory example, we analyse a synthetic dataset where we know all influencing factors.
This dataset is the results of a script generating fibers with characteristics based on a deterministic model determined by the input parameters.
We use the random sequential adsorption method, originally introduced by Widom \cite{widom_random_1966}, to add tubes of constant radius, resembling curved fibers, in a volume of given size.
The length and radius of the fibers are modeled as functions based on the hyperbolic tangent, their values are therefore computed from a numerical integration of a (shifted and scaled) version of \autoref{equ:gaussian}.
One parameter (param1) was modeled to influence the fiber length, another (param2) to influence the fiber diameter.

\begin{equation}
    f(x) = a + b \tanh{c (x+d) }
\end{equation}

Values for $a$, $b$, $c$ and $d$ are chosen so as to shift and scale the resulting output values into ranges typical for real fiber datasets, specifically, a=215, b=15, c=5, d=-0.5 for length, and a=7, b=0.5, c=8, d=-0.3 for diameter.
Values for the input parameters are determined by a pseudo-random number generator in a range from 0 to 1, initialized to a standardized seed, meaning that two runs of the script with the same seed and the same input parameters produce exactly the same result.
We employ the sampling methods from \autoref{sec:sampling} on this script for generating synthetic fibers, to test whether our methods would correctly determine the variations which we predetermined.
That is, the goal of this example is to illustrate the analysis with our methods, and to show that our visualization methods correctly determine the amount of influence of input parameters on the different output characteristics and the respective location in the parameter space.



As shown in \autoref{fig:overview}(a), the in-out-matrix correctly indicates the influence of param1 on the StraightLength characteristic, and that of param2 on the Diameter characteristic. The change in length over the collection of results is also clearly visible through the yellow areas at top and bottom of each fiber in \autoref{fig:overview}(a) (yellow indicating areas covered by few results only). On close inspection, the changes in diameter are also visible as yellow halos around each fiber. The length and diameter changes are uniformly distributed across all fibers, just as they were modeled. In the constellation plots, three results are selected \autoref{fig:constellation}. The result highlighted in orange is very close to the result highlighted in green in the MDS plot. When inspecting it in the parameter space plot, it becomes clear that this is due to their basically identical values for param2. The third selected result in contrast has different values for both input parameters and therefore also is quite far apart in the dissimilarity space.

By inspecting single fibers from these results, as shown in \autoref{fig:detail}(b), we can easily see the length increasing with increasing param1 values.
The difference in diameter for the violet result to the green reference is also clearly visible through colored points over the whole length on its right side.
The parameter influence view indicates the areas in the input parameter range with the highest change through the line plots, as shown in \autoref{fig:detail}(b).
For param1 influencing StraightLength, this is reflected as a broad curve spanning the whole parameter range, with a peak at approximately 0.6.
For param2 and Diameter, the line plot shows a much more pronounced peak at approximately 0.3. This concisely reflects the distributions that were used to generate the length and diameter characteristics. 

The small plateau before the peak for param1, as well as the much higher curve for param2's effect on Diameter than that of param1 on StraightLength can be explained by the fact that the distribution difference measure was used for computing these variations.
As for param1 and its effect on length, the individual length distributions are very broad, as visible in the histogram chart above.
This leads to some overlapping bins with similar frequencies, and the plateau is an effect of the distribution discretization.
With the diameter, the situation is reversed - all fibers in a result have basically the same diameter, and therefore fall into the same histogram bin. Changes that shift the diameter to fall into another distribution bin lead to a completely different distribution.
To overcome this limitation of the distribution difference, there is also the possibility to employ a difference measure that computes average distances between the results on a pairwise match basis, as explained in \autoref{sec:sensitivity-measures}.


\section{Discussion and Future Work}
\label{sec:discussion}

As an immediate first step towards further evaluation, to be addressed in the next revision of this document, we will apply the methods proposed in this paper to case studies of exploring the sensitivity of two parametric fiber reconstruction algorithms.

Regarding sampling, to improve the coverage of the high-dimensional parameter spaces, quasi-random sampling methods utilizing low discrepancy sequences could be used \cite{Nair:1999:Image}.
Additionally, in some circumstances it might be helpful for users to be able to perform additional sampling in regions of specific interest.
In our case studies so far, a sufficient exploration of the space of interesting parameter combinations never required more than one re-sampling. 
Adding an optional re-sampling would also most likely add considerable complexity to the implementation of our methods.
For situations where an adaptive refinement is required,
the progressive latin hypercube sampling strategy proposed by Sheikoleslami and Razavi \cite{Sheikholeslami:2017:ProgressiveLHS} seems promising.

If the analysis of larger collections of results or of results with more objects is required,
there is plenty of room for optimizing the algorithms used in the preprocessing step for speed.
Using octrees for spatial subdivisioning could be used to speed up the dissimilarity computations.
For the projection from fibers to voxel space, GPU acceleration could provide significant speedup.


\ifCLASSOPTIONcompsoc
  \section*{Acknowledgments}
\else
  \section*{Acknowledgment}
\fi

The research leading to these results has received funding from the Austrian Research Promotion Agency (FFG) within the program line "TAKE OFF", FFG grant no. 874540 "BeyondInspection", and by research subsidies granted by the government of Upper Austria in the course of its "X-Pro"project. Initial studies were received funding from the Research Foundation Flanders(FWO) and the Austrian Science Fund (FWF) under the grant numbers G0F9117N and I3261-N36 "Quantitative X-ray tomography of advanced polymer composites" respectively.

\ifCLASSOPTIONcaptionsoff
  \newpage
\fi



\bibliographystyle{IEEEtran}
\bibliography{main.bib}

\end{document}